\documentclass[final]{aipproc}
\usepackage[utf8]{inputenc}

\layoutstyle{6x9}

\begin{document}

\title{Reverse Engineering Quantum Field Theory}

\classification{03.65.Ta, 03.65.Ca, 11.10.Cd}
\keywords{foundations of quantum theory, quantum field theory, general boundary formulation}

\author{Robert Oeckl}{address={Centro de Ciencias Matemáticas,\\ Universidad Nacional Autónoma de México,\\ C.P.~58190, Morelia, Michoacán, Mexico}}

\begin{abstract}
An approach to the foundations of quantum theory is advertised that proceeds by ``reverse engineering'' quantum field theory. As a concrete instance of this approach, the general boundary formulation of quantum theory is outlined.
\end{abstract}

\maketitle

\begin{quote}
``That `sharp time' is an anomaly in Q.M.\ and that besides, so to speak independently of that, the special role of time poses a serious obstacle to adapting Q.M.\ to the relativity principle, is something that in recent years I have brought up again and again, unfortunately without being able to make the shadow of a useful counterproposal.''
E.~Schrödinger, 1935, \cite{Sch:katze}.
\end{quote}

The rules of quantum theory, both mathematical and physical have been fixed essentially in the late 20's and early 30's of the previous century.
Since then, we have learned a lot more about nature and our understanding of the foundations of physics has thoroughly changed. Quantum field theory has become the main tool for describing nature at the most basic level accessible to us. Special relativity has become an indispensable and fundamental ingredient of this description. At large scales, general relativity is at the basis of our understanding of nature.
Nevertheless, when we consider the foundations of quantum theory itself, we retreat to a picture that was developed as an analogy to non-relativistic classical physics and is heavily imprinted with this legacy.

In this picture, a Hilbert space is associated to any ``system''. The ``states'' of the system are the elements of this Hilbert space. A Hamiltonian operator describes the time-evolution of these states in the absence of measurements. A measurement also corresponds to an operator on this Hilbert space. A measurement is \emph{instantaneous} and apart from producing a result, \emph{modifies} a state (in a partially random way). The \emph{temporal} composition of measurements is encoded in the non-commutative product of the corresponding operators. Clearly, a predetermined notion of time is essential to making sense of any quantum theory in this picture, which we shall refer to as the \emph{standard formulation of quantum theory}.\footnote{There are axiomatic approaches to quantum field theory involving structures different from the ones described. For example, in algebraic quantum field theory observable algebras play a more central role and Hilbert spaces appear as a derived concept \cite{Haa:lqp}. Nevertheless, for the purposes of the present contribution, these axiomatic systems and their physical interpretation are sufficiently similar to be subsumed under the standard formulation.}

The situation is very different in classical physics. Here, a (perfect) measurement does not affect the measured system and the temporal order of measurements is unimportant for the outcomes. There is also no fundamental reason to conceptualize a measurement as being instantaneous. There is thus no intrinsic need for a predetermined notion of time and hence no conceptual barrier to implementing descriptions of special relativistic or even general relativistic physics.

Special relativity does not provide a unique predetermined notion of time. However, it provides various such notions, one per inertial frame, and they are tightly related. This relation makes it possible to put special relativistic theories within the standard formulation of quantum theory. To this end, observables are localized in spacetime. Then, the operators corresponding to observables are required to commute whenever the observables are relatively spacelike located. This ensures that potential ordering ambiguities due to the different notions of time associated with different choices of inertial frame are precisely avoided.
In this way, even quantum field theory with special relativity as a core ingredient, is based on the standard formulation of quantum theory. At least, this is the way we usually think about it and present it in text books. It is, however, an uneasy marriage. As already mentioned, observables need to be equipped with explicit labels indicating their spacetime location. This is an ingredient alien to the original spirit of the standard formulation. Moreover, and this will lead us to the main theme of this contribution, the structures of the standard formulation are not necessarily the operationally relevant ones in quantum field theory.

In the paper
cited at the beginning of this contribution it becomes evident that Schrödinger, in the early days of quantum mechanics, was not only quite aware of the special role of time in the standard formulation, but even considered the latter provisional in light of this deficiency. (What he called Q.M.\ or quantum mechanics is what we call here the standard formulation.) It is fair to assume that at least a minority of other physicists at the time shared these doubts. In spite of the concerns, the original formulation prevailed and is rarely questioned today. One reason is of course that it was shown to be compatible with special relativity after all. Moreover, it is seen as part of the extraordinary empirical success story of quantum field theory culminating in the Standard Model of Elementary Particle Physics.

There is another reason for the reluctance to reconsider the foundations of quantum theory. In the mind of many physicists (the author included) quantum theory works a bit like a black box. You know exactly how to use it, but you don't understand what is inside it. It works extraordinarily well and allows you to make extremely accurate predictions. Understandably, you are reluctant to tamper with it, for fear of breaking it.
But the benefits of a reconsideration could be considerable. Schrödinger's concern that \emph{``the special role of time poses a serious obstacle to adapting Q.M.\ to the relativity principle''} remains spot on even today, at least if by \emph{``relativity principle''} we understand general relativity. Thus, overcoming this special role would be particularly promising. Moreover, a deeper understanding of what quantum theory ``really is'', is an important objective in its own right.

So, how should one go about this? First of all, we have an empirically hugely successful description of most of fundamental physics known today, based on quantum field theory. It would be foolish to ignore this and pretend that we have any chance of success if we tried to reinvent physics from scratch. Instead, we shall take seriously the predictive power of quantum field theory and the relevant machinery behind it. But at the same time, we shall question the text book narrative of its foundation on the standard formulation of quantum theory. Rather, we shall look for a new formulation of quantum theory on which the operationally relevant parts of quantum field theory rest more naturally. Our main source for clues about this new formulation shall be quantum field theory itself. We shall refer to this programme as \emph{reverse engineering quantum field theory}.
An apparent disadvantage of this approach is its ambiguity. As in any inverse problem there is probably not a unique solution. One might hope to compensate for this by adding further guiding principles. General covariance would certainly be highly desirable. The approach also has a very compelling advantage. It avoids the danger of ``breaking'' quantum theory. Indeed, it is \emph{designed} to preserve known quantum physics.

The \emph{general boundary formulation of quantum theory (GBF)} \cite{Oe:gbqft} is an approach to the foundations of quantum theory precisely in the spirit of this programme. We shall spend the rest of this contribution focused on it, explaining in particular in which sense it amounts to a reverse engineering of quantum field theory.

It is clear that precisely points of tension between quantum field theory and the standard formulation are promising starting points for our search. We shall focus here on the following structural features of operational quantum field theory:
\begin{description}
\item[The Feynman path integral.] This turns out to be much more
 suitable to describe the dynamics of quantum field theory than
  Hamiltonian or time-evolution operators.
\item[Crossing symmetry.] This property of the S-matrix is completely
 unmotivated from the point of view of the standard formulation.
\item[The time-ordered product of fields.] This rather than the operator
   product is the relevant structure to extract physical
  predictions.
\end{description}

The path integral was originally conceived by Feynman as a means to capture the dynamics of a quantum system, in particular a non-relativistic one \cite{Fey:stnrqm}. The time-evolution from an initial time $t_1$ to a final time $t_2$ is obtained as a certain integral over all possible paths of participating particles in the time interval $[t_1,t_2]$. It is then easy to see that this integral has a required composition property: If we evolve the system from a time $t_1$ to a time $t_2$ and then to a time $t_3$ we obtain the same result as evolving directly from time $t_1$ to time $t_3$. Indeed, taking the product of the path integral corresponding to $[t_1,t_2]$ with that corresponding to $[t_2,t_3]$ and summing over all intermediate configurations at $t_2$ recovers the path integral corresponding to $[t_1,t_3]$.

Subsequently, the Feynman path integral has become an ubiquitous tool in quantum field theory. There, rather than over particle trajectories it is an integral over spacetime field configurations. Correspondingly, its composition property is generalized from temporal to spacetime composition: Let $M_1$ and $M_2$ be adjacent spacetime regions that together from the joint spacetime region $M$. Then, taking the product of the path integrals in $M_1$ and in $M_2$ and summing over all intermediate field configurations on the joint boundary yields the path integral for $M$.

Abstracting these and other remarkable properties of the Feynman path integral, lead in the 80's of the past century to the development of \emph{topological quantum field theory (TQFT)} \cite{Ati:tqft}, initiated by E.~Witten. Through the participation of mathematicians such as G.~Segal and M.~Atiyah this developed into a new branch of modern algebraic topology with connections to knot theory, low dimensional topology, categorical algebra and others. In brief, a TQFT is a collection of geometric data and an associated collection of algebraic data. The geometric data consists of a collection of manifolds of dimension $d$, that we shall call \emph{regions}, together with a collection of manifolds of dimension $d-1$, that we shall call \emph{hypersurfaces}, and which arise as boundaries of the regions or pieces of such boundaries. To each hypersurface one associates a complex vector space. To each region one associates a linear \emph{amplitude} map from the vector space on its boundary to the complex numbers. These structures satisfy a number of axioms, making them fit together in a nice way. One such axiom is that the vector space associated to a hypersurface with several components is the tensor product of the vector spaces associated to the individual components. Another axiom, called the \emph{gluing axiom}, says that we may combine two adjacent regions to a new one by gluing along a common boundary hypersurface. The amplitude map associated to the new region is then obtained by taking the product of the amplitude maps for the original regions and summing over a basis of the vector space on the common hypersurface.

Thinking of the regions and hypersurfaces in a TQFT as representing pieces of spacetime clarifies its motivation from quantum field theory and the path integral. In particular, the gluing axiom abstracts precisely the composition property of the path integral. In this context, the vector space associated to a hypersurface of constant time would be the Hilbert space of states of the system (at that time). The amplitude map associated to a time-interval region would, for a given tensor product of initial and final state, be the transition amplitude from the initial to the final state.
One of the interesting features of TQFT is that the regions and hypersurfaces involved need not carry any additional structure besides a topology. 
This means that the formalism as such is compatible from the outset with general relativistic principles.

The starting point of the GBF consists of taking seriously a suitable TQFT as a mathematical codification of fundamental quantum physics \cite{Oe:boundary}. A priori it is not at all clear that this can be successful. One peculiar consequence of this step the following. Consider again a transition amplitude as described above. The fact that an element of the boundary Hilbert space decomposes into (a linear combination of) initial and final state pairs is due to the decomposition of the boundary hypersurface itself into and initial and final component. (Recall the tensor product axiom.) However, consider a generic region whose boundary does not decompose into components. There would then not be a geometric way to distinguish, say, incoming and outgoing particles. Therefore, the amplitude for a multi-particle state needs to be essentially independent of which particle we consider as incoming and which as outgoing. This is quite a striking property and it is completely unmotivated from the standard formulation. However, it is indeed true in quantum field theory and it is called \emph{crossing symmetry} there.
Crossing symmetry was indeed one of the original motivations for the GBF \cite{Oe:catandclock} and its explicit emergence from a setting with a connected boundary in the GBF was later shown \cite{CoOe:smatrixgbf}.

A crucial question is how measurable quantities can be extracted from these mathematical objects. For time-interval regions and the associated transition amplitudes this is clear (taking the modulus square etc.). For general spacetime regions this is not obvious and cannot be derived from the standard formulation. However, a coherent and compelling probability interpretation is possible and constitutes probably the most important achievement of the GBF to date \cite{Oe:gbqft,Oe:probgbf}. Crucially, it reduces to the probability interpretation of the standard formulation where applicable. We shall not discuss it here, however, since known features of quantum field theory do not seem to play any role.

Instead, we proceed to the consideration of \emph{observables}. As already mentioned, in quantum field theory these are labeled by spacetime locations. Moreover, when composing the operators corresponding to such observables the only operationally meaningful composition is by ordering according the the temporal labels. This is called the \emph{time-ordered product}. The original non-commutative operator product never directly enters the calculation of measurable quantities in quantum field theory. This is quite striking and throws into doubt the naturalness of the formalization of observables as operators in the first place. On the other hand, the time-ordered product naturally arises from the Feynman path integral. Inserting a classical observable, i.e., a real function of field configurations, into the path integral yields the quantized observable as a generalized amplitude. Moreover, if the classical observable is a product of localized observables, the path integral yields precisely the time-ordered product of the individual quantized observables. (The non-relativistic version of this was understood already by Feynman in his original article on the path integral \cite{Fey:stnrqm}.)

All this suggests to formalize observables as located in spacetime regions and giving rise to linear maps from the region's boundary Hilbert space to complex numbers, similar to amplitudes \cite{Oe:obsgbf}. The natural composition of observables is then given by the gluing of their carrier regions. Also, the probability interpretation mentioned above extends to these observables and yields an appropriate notion of \emph{expectation value}, reducing the the usual one when applicable. That this formalization of observables indeed nicely subsumes the concept of observable in quantum field field theory was shown in \cite{Oe:feynobs}.

While the GBF appears to be a promising path towards new foundations in the spirit of the ``reverse engineering'' programme, it remains to be demonstrated that it can be made solid enough for quantum field theory in all of its complexity to rest on it.

\begin{theacknowledgments}
This work was supported in part by UNAM–DGAPA–PAPIIT through project
grant IN100212.
\end{theacknowledgments}

\bibliographystyle{aipproc}
\bibliography{stdrefsb}

\end{document}